\documentclass[prl,reprint,superscriptaddress]{revtex4-1}
\usepackage{graphicx}

\begin{document}

\title{Optical and spin properties of a single praseodymium ion in a crystal}

\author{Kangwei Xia}
\email[]{k.xia@physik.uni-stuttgart.de}
\affiliation{3. Physikalisches Institut, Universit$\mathrm{\ddot{a}}$t Stuttgart, 70550 Stuttgart, Germany}
\affiliation{Department of Physics, The Chinese University of Hong Kong, Shatin, New Territories, Hong Kong, China}
\author{Roman Kolesov}
\affiliation{3. Physikalisches Institut, Universit$\mathrm{\ddot{a}}$t Stuttgart, 70550 Stuttgart, Germany}
\author{Ya Wang}
\email[]{ywustc@ustc.edu.cn}
\affiliation{Hefei National Laboratory for Physics Sciences at Microscale and Department of Modern Physics, University of Science and Technology of China, Hefei, 230026, China}
\author{Petr Siyushev}
\affiliation{Institute for Quantum Optics and Center for Integrated Quantum Science and Technology (IQst), Universit$\mathrm{\ddot{a}}$t Ulm, 89081 Germany}
\author{Thomas Kornher}
\affiliation{3. Physikalisches Institut, Universit$\mathrm{\ddot{a}}$t Stuttgart, 70550 Stuttgart, Germany}
\author{Rolf Reuter}
\affiliation{3. Physikalisches Institut, Universit$\mathrm{\ddot{a}}$t Stuttgart, 70550 Stuttgart, Germany}
\author{Sen Yang}
\affiliation{3. Physikalisches Institut, Universit$\mathrm{\ddot{a}}$t Stuttgart, 70550 Stuttgart, Germany}
\affiliation{Department of Physics, The Chinese University of Hong Kong, Shatin, New Territories, Hong Kong, China}
\author{J\"{o}rg Wrachtrup}
\affiliation{3. Physikalisches Institut, Universit$\mathrm{\ddot{a}}$t Stuttgart, 70550 Stuttgart, Germany}
\affiliation{Max Planck Institute for Solid State Research, Heisenbergstraße 1, 70569 Stuttgart, Germany}


\date{\today}

\begin{abstract}
The investigation of single atoms in solids, with both optical and nuclear spin access is of particularly interest with applications ranging from nanoscale sensing to quantum computation. Here, we study the optical and spin properties of single praseodymium ions in an yttrium aluminum garnet (YAG) crystal at cryogenic temperature. The single nuclear spin of  single praseodymium ions is detected through a background-free optical upconverting readout technique. Single ions show stable photoluminescence (PLE) with spectrally resolved hyperfine splitting of the praseodymium ground state. Based on this measurement, optical Rabi and optically detected magnetic resonance (ODMR) measurements are performed to study the spin coherence properties. Our results reveal that the spin coherence time of single praseodymium nuclear spins is  limited by the strong spin phonon coupling at experimental temperature.

\end{abstract}

\maketitle

Optical detection and coherent control of single quantum objects in solids is essential to various fields ranging from fundamental physics to  quantum information technologies~\cite{ladd2010, awschalom2013, gao2015}. Among these single photon emitters, rare earth ions embedded in crystals are attracting increasing attention as they simultaneously show narrow optical transitions~\cite{thorpe2011, equall1994, Riedmatten2008} and long spin coherence times~\cite{konz2003, Saglamyurek2015, zhong2017} as well as on-chip photonics~\cite{faraon2015, ding2016, sinclair2016}. A single rare earth ion having nuclear spin, which allows optical addressing and control, is particularly interesting as it potentially combines with ultra-long coherence time with fast optical control. However, it still remains challenging to coherently address single rare earth ions with nuclear spins~\cite{kolesov2012, utikal2014, nakamura2014}.

Up to now, only single electron spins of trivalent cerium ions in YAG have been optically detected, initialized, readout and coherently controlled~\cite{kolesov2013, siyushev2014, xia2015}. Due to the strong coupling to the surrounding spin baths,  single Ce electron spin qubits loses their coherence on a sub-microsecond time scale. Although it is extended to millisecond by dynamical decoupling technique~\cite{siyushev2014}, the coherence time is still far below that of rare earth ion nuclear spins. Excellent examples are trivalent Pr nuclear spins with one minute coherence time~\cite{heinze2013} and Eu:YSO nuclear spins with coherence times up to six hours~\cite{zhong2015}.
However, detection and coherent control  of these two rare-earth ion single nuclear spin is hampered by their low fluorescence intensity.

In this work, we report the optical and spin properties of a single nuclear spin of praseodymium ion in YAG crystal at cryogenic temperature. In particular, we observe the spectrally well resolved ground state hyperfine of praseodymium ions and demonstrate the initialization and readout of single nuclear spin through optical control. Moreover, we also performed optical Rabi and ODMR measurements to demonstrate optical control and radio frequency(RF) control capabilities of single praseodymium ions.
The experiments we demonstrated open the door to future implementation of more sophisticated techniques, such as all-optical control and dynamical decoupling control~\cite{yale2016}.

\begin{figure}[h!]
\includegraphics[width=3.5in]{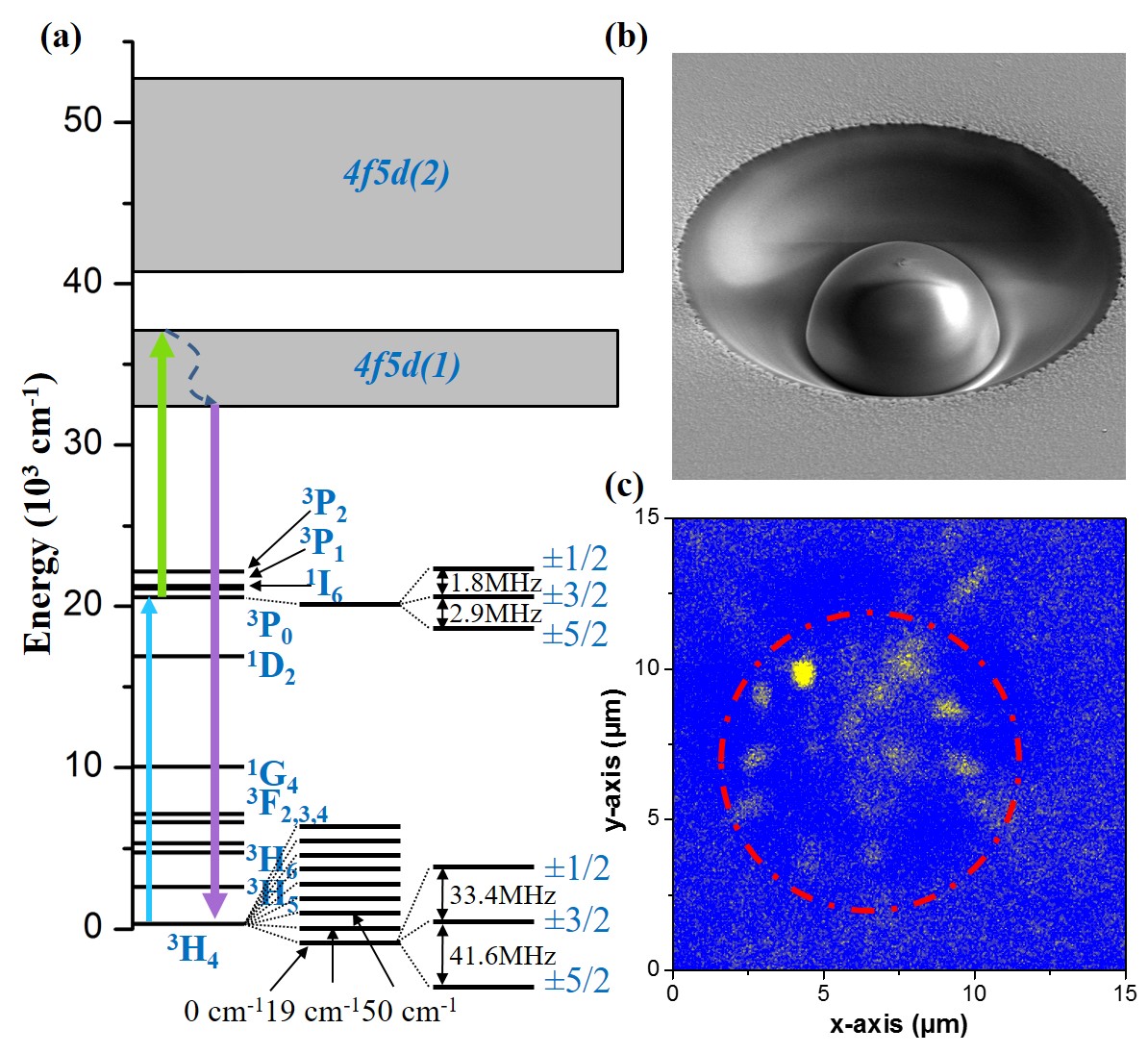}%
\caption{ Optical detection of single praseodymium ions in YAG crystal. (a) Energy level of Pr:YAG.  Upconversion process of Pr:YAG indicated by the blue and green arrows. The Pr$^{3+}$ ion is first excited from $\mathrm{^3H_4(1)}$ to $\mathrm{^3P_0}$ state by a 487~nm wavelength laser. It is then promoted to the $4f5d$ band by absorbing the second photon of 532~nm wavelength. (b) Scanning electron microscopy (SEM) image of a SIL fabricated by focused ion beam milling. (c) Upconversion image of individual Pr ions underneath a SIL.  \label{Fig1}}
\end{figure}

The energy level structure of Pr:YAG is presented in Fig.~\ref{Fig1}. The ground state $\mathrm{^3H_4(1)}$ is located in the 4$f$ shell, and is composed of three doubly degenerate hyperfine sublevels with energy level splitting of 33.4 and 41.6~MHz. In rare earth ions, the intra-4$f$-shell transitions are efficiently screened by closed outer lying 5$s$ and 5$p$ shells. This screening is responsible for  narrow optical $4f^2 \leftrightarrow 4f^2$ transitions~\cite{kim1991}. The $\mathrm{^3P_0}$ state shows a lifetime of 8~$\mu$s~\cite{cheung1994}. The long lifetime of the excited states and their resulting weak fluorescence challenges the detection of single rare earth ions directly through $4f^2 \leftrightarrow 4f^2$ transitions, however, has recently been detected~\cite{utikal2014, nakamura2014, Eichhammer}.

One can circumvent the long lifetime $\mathrm{^3P_0}$ by further exciting the ion to the $4f5d$ state. Compared to $\mathrm{^3P_0}$ the $4f5d$ state shows a much shorter lifetime of 18~ns~\cite{gayen1992, ganem1992}. Thus, the emission  rate of single ions can be largely enhanced by two-photon upconversion. Previously, we used this upconversion method for the first optical detection of single Pr ions at room temperature~\cite{kolesov2012}. In order to get  access to the nuclear spin degrees of freedom, the same detection technique is applied at cryogenic temperatures in this work.

Single Pr$^{3+}$ ions are detected through an upconversion process $\mathrm{^3H_4(1) \rightarrow^3P_0}\rightarrow 4f5d$, as shown in Fig.~\ref{Fig1}(a)). A broadband diode laser with wavelength at 487~nm is applied to excite the $\mathrm{^3H_4\rightarrow^3P_0}$ transition. Another 532~nm laser is applied simultaneously to promote the Pr$^{3+}$ ion further to the $4f5d$ band ($\mathrm{^3P_0}\rightarrow 4f5d$). The emitted photons  collected by a 0.85 N. A. objective lens are detected by a photomultiplier tube (PMT) in a spectral range between 300-400 nm. Figure~\ref{Fig1}(b) shows the SEM image of a solid immersion lens fabricated on the surface of the studied YAG crystal in order to enhance  photon collection efficiency and spatial resolution~\cite{Jamali2014}. Figure~\ref{Fig1}(c) displays the  laser scanning fluorescence image of Pr ions underneath a SIL, where bright spots represent individual Pr ions.

\begin{figure}[h]
\includegraphics[width=3.2in]{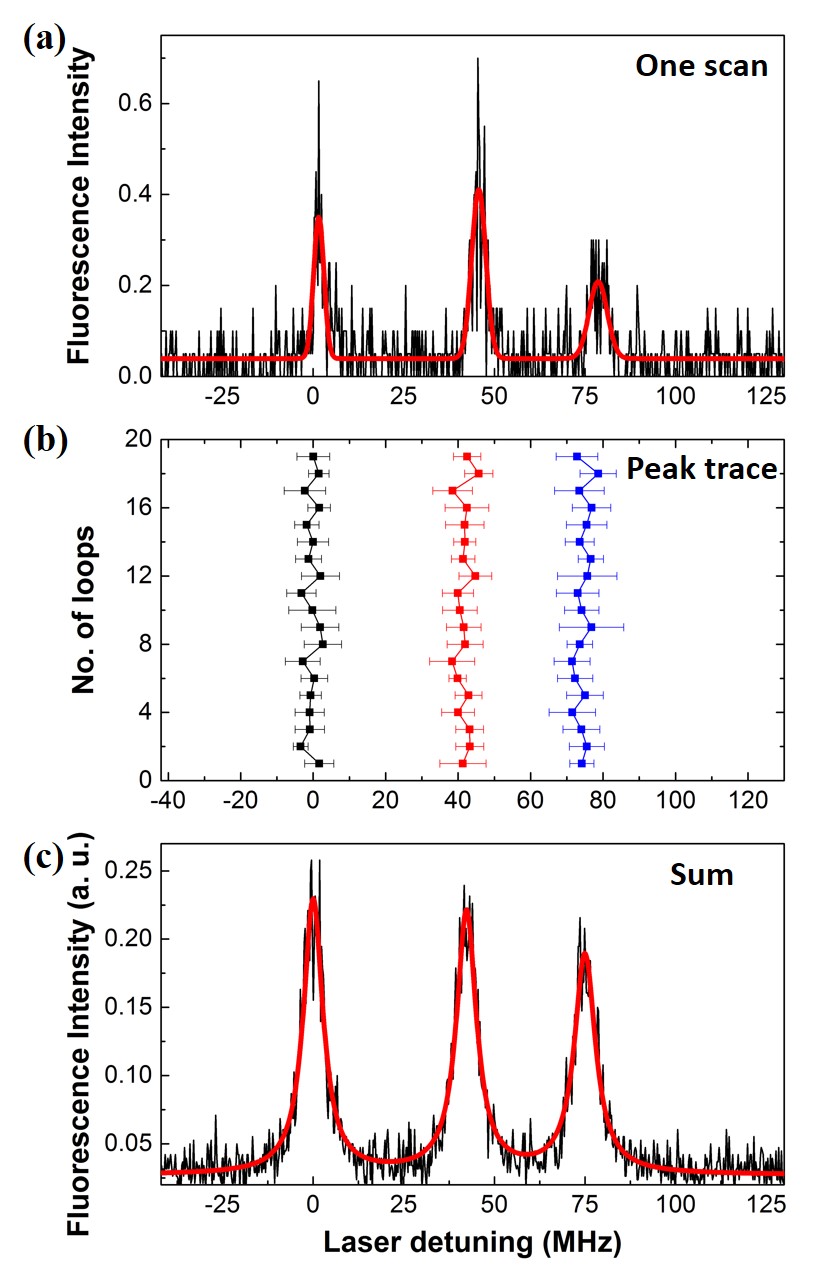}%
\caption{Upconverted PLE of a single Pr ion in YAG. (a) Single sweep of upconverted PLE spectrum. (b) Plot of successive sweeping. Square spots represent center peak frequency. Error bar represents the width of the peak. (c) Sum of upconverted PLE spectra of a single Pr ion, indicating $\sim2\pi\times$6~MHz optical transition linewidth. \label{Fig2}}
\end{figure}

The spectral properties of single Pr ions were investigated by PLE measurements. To this end, a narrow linewidth single mode laser (Toptica Photonics, DL Pro) working at 487~nm used for narrow band optical excitation. To prevent power broadening of the transitions, the laser was kept at low power levels ($\sim$5~$\mathrm{\mu}$W). While the 487~nm laser was swept through the resonant $\mathrm{^3H_4\rightarrow ^3P_0}$ transition, the 532~nm laser was constantly illuminating the sample completing the second upconversion step. By monitoring the fluorescence during frequency sweeping of the single mode laser, a single sweep PLE spectra is obtained as shown in Fig.~\ref{Fig2}(a). The spectrum shows a background-free signal as the upconversion readout method efficiently filters out the noise.  From the spectrum one can see three well-resolved peaks with frequency differences of 32.7~MHz and 42.3~MHz, respectively. These peaks correspond to the hyperfine splittings of the ground state $\mathrm{^3H_4}$(1). However the hyperfine structure of the excited state $\mathrm{^3P_0}$ is not resolved owing to the linewidth of $\mathrm{2\pi\times5~MHz}$.
By monitoring the peak frequency in the successive sweeps as shown in Fig.~\ref{Fig2}(b), we obtain a spectral diffusion of single Pr$^{3+}$ ions within $\pm$3~MHz.
Fig.~\ref{Fig2}(c) is an average of the spectra of the successive sweeps in Fig.~\ref{Fig2}(b). The total linewidth is $\sim\mathrm{2\pi\times6~MHz}$ which includes both, spectral diffusion and intrinsic linewidth.

Figure 3 shows the optical coherent control of a single Pr ions. We applied a resonant laser with varying pulse length to drive it oscillating between the $\mathrm{^3H_4}$ and $\mathrm{^3P_0}$ states. As a result of its very weak oscillation strength, we observe an oscillation frequency of only $\mathrm{2\pi \times 51}$~MHz by applying a laser power 0.4 mW. This value is several orders of magnitude smaller than that of other atom-like systems, such as NV centers~\cite{Batalov2008, Golter2014} and quantum dots~\cite{Stievater2001, Kamada2001}.

\begin{figure}[h]
\includegraphics[width=3.4in]{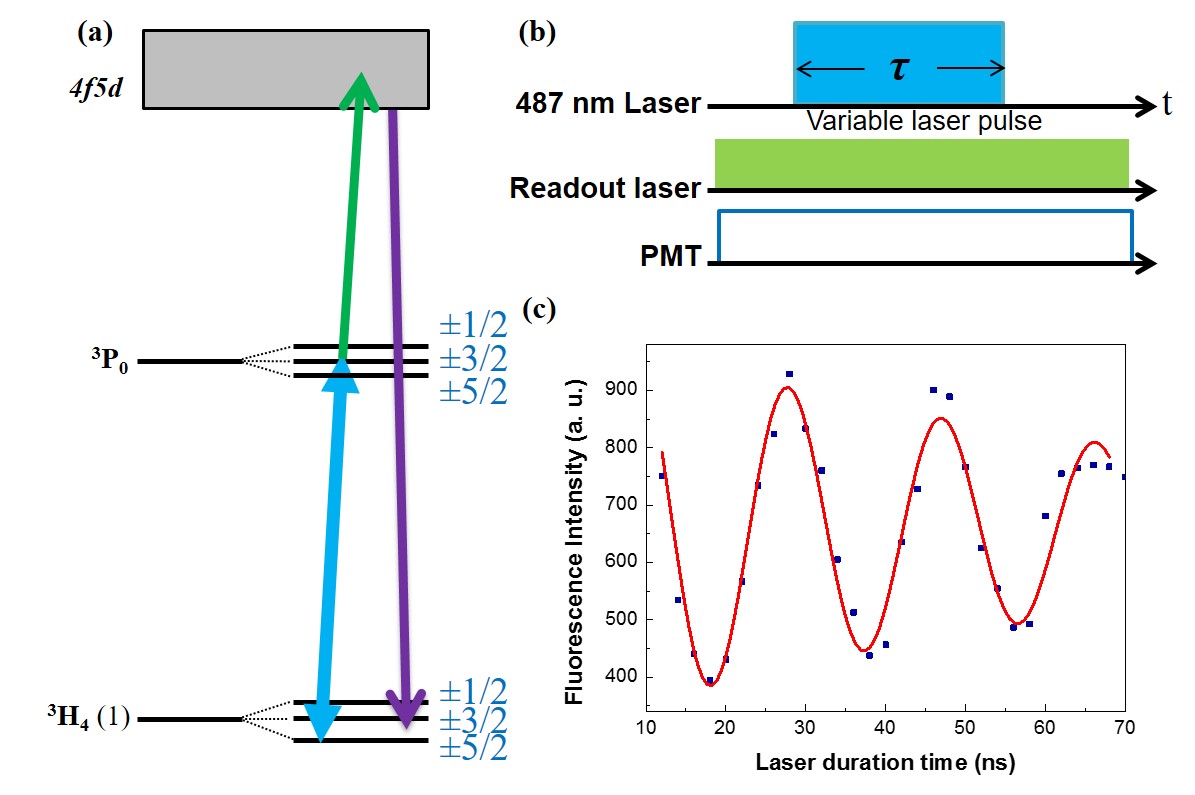}%
\caption{Optical Rabi of single Pr ion. The laser intensity of the 4$f \leftrightarrow 4f$ transition is 0.4 mW. (a) Scheme of optical Rabi oscillation of single Pr ions in YAG (b) Pulse sequences: a green laser was constantly applied to accomplish the upconversion process. The duration time of the blue laser was various. (c) Optical Rabi oscillation of single Pr ions in YAG.  \label{Fig4}}
\end{figure}

In addition to optical control, the ground state hyperfine splitting resolved in PLE spectrum enables resonant optical initialization and readout of single Pr nuclear spins. Thus, hyperfine resolved control and investigation of spin coherence properties by means of ODMR measurements are feasible.

\begin{figure}[h]
\includegraphics[width=3.4in]{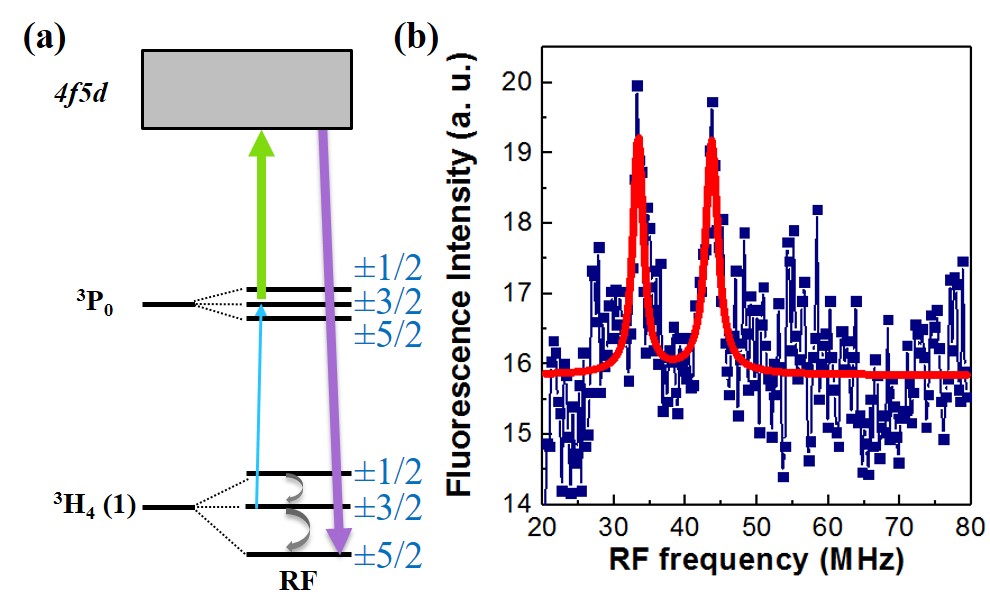}%
\caption{ Upconverted ODMR of a single Pr ion in YAG. (a) Energy diagram of the ODMR, the single mode narrow linewidth laser is on resonance with either of an optical transitions ($\mathrm{^3H_4 (\pm 3/2)\rightarrow ^3P_0 (\pm 3/2)}$). The 532 nm green laser is added to promote single Pr ions further to $4f5d$ band, while the RF frequency is swept. (b) Upconverted ODMR of a single Pr ions in YAG crystal.  \label{Fig3}}
\end{figure}

Figure~\ref{Fig3}(a) shows the ODMR measurement by applying the single mode laser  on resonance with the optical transition $\mathrm{^3H_4 (\pm 3/2)\rightarrow ^3P_0 (\pm 3/2)}$  and a RF source  simultaneously to induce spin transitions. The laser power was set to 5 $\mu$w, corresponding to a excitation rate of $2\pi \times 5.7~$MHz according to the optical Rabi measurement in Fig.3. This rate is much faster than the spontaneous rate of $\mathrm{^3P_0}$ state, ensuring the $\mathrm{^3H_4(3/2)}$ spin state is totally depleted~\cite{Ham1997, Turukhin2001}. By sweeping the RF frequency,  upconverted ODMR spectra were acquired, which are shown in Fig.~\ref{Fig3}(b). The figure shows transition peaks at frequency 33.4 and 43.7~MHz, in good agreement with  PLE measurement.
The linewidth of the ODMR spectrum is $\sim2\pi\times$1.5~MHz, indicating that the coherence time of single Pr nuclear spin is one order of magnitude longer than that of single Ce electron spins in the same crystal~\cite{siyushev2014}, but far shorter than the expected values by taking into account the three orders of magnitude gyromagnetic ratio difference between the electron and nuclear spins. To understand this discrepancy we note that the baseline in Fig.~\ref{Fig3}(b) has an obvious fluorescence intensity. Since the upconversion readout under low power excitation is background free as shown in Fig.~\ref{Fig2}(a), the observed finite fluorescence intensity indicates the probability of that single Pr ions have a certain probability for staying in state $\pm \frac{3}{2}$. We thus attribute the non-zero baseline and the linewidth broadening to the spin-lattice relaxation. A fast spin-lattice relaxation rate, which is comparable to the laser pumping rate, causes single Pr ions relaxed back to the state $\pm \frac{3}{2}$ and also broadens the magnetic resonance peak.

To understand the ODMR process more precisely, we describe the ODMR measurement through optical transition by a modified three-level Bloch equation:
\begin{eqnarray*}
\frac{\mbox{d}\rho_{aa}}{\mbox{d}t} &=& i\Omega (\rho_{ab}-\rho_{ba}) - \frac{\rho_{aa}}{2T_1} + \frac{\rho_{bb}}{2T_1} + \frac{\Gamma}{2}\rho_{bb}\\
\frac{\mbox{d}\rho_{bb}}{\mbox{d}t} &=& -i\Omega (\rho_{ab}-\rho_{ba}) - \frac{\rho_{bb}}{T_1} + \frac{\rho_{aa}+\rho_{cc}}{2T_1} - \Gamma{\rho_{bb}}\\
\frac{\mbox{d}\rho_{cc}}{\mbox{d}t} &=& - \frac{\rho_{cc}}{2T_1} + \frac{\rho_{bb}}{2T_1} + \frac{\Gamma}{2}\rho_{bb}\\
\frac{\mbox{d}\rho_{ab}}{\mbox{d}t} &=& i\Delta \rho_{ab} + i \Omega (\rho_{aa}-\rho_{bb}) - \frac{\rho_{ab}}{T_2}
\end{eqnarray*}
where the symbols $a,b,c$ represent the sublevels $\pm 1/2$, $\pm 3/2$, and $\pm 5/2$ respectively. $\Delta$ is the detuning of the RF frequency, $\Omega$ is the RF induced Rabi frequency, $T_1$ and $T_2$ are the spin relaxation times. $\Gamma$ is the population redistribution rate determined by the excitation-emission cycling rate under optical pumping. In this model, the population redistribution rates from the sublevels $\mathrm{\pm3/2}$ to $\mathrm{\pm1/2}$ and $\mathrm{\pm5/2}$ are considered to be identical as indicated by identical ODMR peak intensity observed in Fig.3(b). In addition we assume the spin relaxation times are the same for all spin transitions.

The optical excitation is a two-step transition with the combination of $\mathrm{^3H_4\rightarrow^3P_0}$ transition and upconverted readout $\mathrm{^3P_0\rightarrow}4f5d$ transition.  Both the oscillation strength and the applied laser power (5~mW) for the transition $\mathrm{^3P_0\rightarrow}4f5d$ is much stronger than for the transition $\mathrm{^3H_4\rightarrow^3P_0}$. The excitation rate is thus dominated by the optical pumping rate of $\mathrm{2\pi\times5.7~MHz}$ between $\mathrm{^3H_4\rightarrow^3P_0}$.
This rate is much slower than the spontaneous emission rate of state 4$f$5$d$ and thus determines the decay rate of $\Gamma=\mathrm{2\pi\times5.7~MHz}$.

The steady solution for $\rho_{bb}$ in the resonant condition ($\Delta = 0,~\rho_{bb\tiny{\mbox{(on)}}}$) and off-resonant condition ($\Delta \gg 0,~\rho_{bb\tiny{\mbox{(off)}}}$) give the relative intensity of the ODMR peak and baseline. The contrast of the ODMR peak defined by:
\begin{equation}
\mbox{Constrast}=\frac{\rho_{bb\tiny{\mbox{(on)}}}-\rho_{bb\tiny{\mbox{(off)}}}}{\rho_{bb\tiny{\mbox{(off)}}}},
\end{equation}

it can be deduced as:

\begin{equation}
\mbox{Constrast} = \frac{4T^{3}_{1}\Gamma\Omega^2}{3+2T_1\Gamma+12T^2_{1}\Omega^2 + 4T^{3}_{1}\Gamma\Omega^2}
\end{equation}
where $T_2 = 2T_{1}$ is assumed here. The RF induced Rabi frequency $\Omega$ is estimated to be $2\pi\times15\pm5$~kHz by taken the waveguide structure and the RF power~\cite{wang2015}. We deduced the contrast to be $\mathrm{14\pm5}\%$. According to these parameters we obtained the $T_1$ time of single Pr ion nuclear spin at 4~K is estimated to be $\mathrm{3.6\pm1.8~\mu}$s.   This fast relaxation time is consistent with the measured ODMR linewidth. 

In rare-earth ions system Orbach relaxation is usually the dominated mechanism for spin-lattice relaxation~\cite{orbach}.
\begin{equation}\label{Orbach}
     T_1 = \frac{1}{c \cdot \Delta}exp(\frac{\Delta}{kT})，
\end{equation}
$T_1$ follows Eq.~(\ref{Orbach}), which is exponentially dependent on the energy difference between the lowest and the second lowest ground states $\Delta$ and temperature $T$. From Fig.~\ref{Fig1} we see that for single Pr ions in YAG $\delta E$ is 19~cm$^{-1}$, which is much smaller than other hosts like YSO with 57~cm$^{-1}$. According to  literature, at 4~K, the nuclear spin lifetime of Pr:YSO is 100~s~\cite{nilsson2004}. If we assume Pr:YAG and Pr:YSO have same pre-factor $c$  here, Pr:YAG then has  $\sim$100~$\mathrm{\mu}$s spin lifetime at the same temperature, which is close to our estimation here. The probable  difference between YSO and YAG hosts is  the pre-factor difference and temperature inaccuracy of the experiment. However, the pre-factor of Pr:YAG is not well known, we applied the pre-factor of Pr:YSO here, which might introduce the difference. Meanwhile, the Orbach relaxation time is exponentially dependent to the reciprocal of the temperature. The inaccurate of the experiment temperature will thus also contribute to the difference.

In conclusion, spectroscopic studying of single Pr ions in YAG has been performed at cryogenic temperature. Single Pr ions in YAG show narrow optical transition linewidth and stable optical transition. However, the intrinsic spin property is dominated by  electron phonon coupling even at 4~K.  It suggests that the full application of single Pr ions should be performed at  lower temperature. Moreover, we can improve the total collection efficiency of the single Pr fluorescence once Pr ions are coupled to photonic devices like nano-rods or photonic cavities. The improved optical property and spin property of single Pr ions in YAG will offer a new opportunity of exploring the long lived nuclear spins.\\

\begin{acknowledgments}

We would like to thank Rainer St\"{o}hr and Nan Zhao for discussions. The work is financially supported by ERC SQUTEC, EU-SIQS SFB TR21 and DFG KO4999/1-1. Ya Wang thanks the supporting from 100-Talent program of CAS.\\

\end{acknowledgments}

\end{document}